\documentclass{PoS}
\RequirePackage{xspace}
\newcommand{\Dbar}{\kern 0.18em\overline{\kern -0.18em D}{}\xspace}

\newcommand{\Dz}{\ensuremath{D^0}\xspace}
\newcommand{\Dzb}{\ensuremath{\Dbar^0}\xspace}
\newcommand{\DzDzb}{\ensuremath{\Dz {\kern -0.16em \Dzb}}\xspace}
\newcommand{\DsP}{\ensuremath{D_S^+}\xspace}
\newcommand{\DsM}{\ensuremath{D_S^-}\xspace}
\newcommand{\DspDsm}{\ensuremath{\DsP {\kern -0.16em \DsM}}\xspace}
\newcommand{\Dp}{\ensuremath{D^+}\xspace}
\newcommand{\Dm}{\ensuremath{D^-}\xspace}
\newcommand{\DpDm}{\ensuremath{\Dp {\kern -0.16em \Dm}}\xspace}

\title{First Results from BESIII}

\ShortTitle{First Results from BESIII}

\author{\speaker{Hai-Bo Li on behalf of the BESIII Collaboration}\\
        Institute of High Energy Physics, Beijing, 100049, China\\
        E-mail: \email{lihb@ihep.ac.cn}}

\abstract{We present the most recent results from the BESIII
experiment. This review covers the studies of charmonium decays,
light hadron spectroscopy and charm physics. Especially, the
prospects for weak decays of charm mesons are addressed at the
BESIII.}

\FullConference{Flavor Physics and CP Violation - FPCP 2010\\
        May 25-29, 2010\\
        Turin, Italy}

\begin{document}

\section{Introduction}

The Beijing Electron Collider has been upgraded (BEPCII) to a
   double ring collider with a design luminosity of $1\times
   10^{33}$cm$^{-2}$s$^{-1}$ at a center-of-mass energy of
   3.78GeV. It is operating between 2.0 and 4.6 GeV in the center
   of mass. The BESIII experiment will be used to study the charm and $\tau$ physics.
   It is foreseen to collect on the order of 10 billion $J/\psi$
events or 3 billion $\psi(2S)$ events per year according to the
designed luminosity.  About 32 million $D\overline{D}$ pairs and
$2.0$ million $D_S \overline{D}_S$ at threshold will be collected
per year~\cite{bes3book,bes3nim}.  In last summer, the peak
luminosity of BEPCII had reached $3.2 \times
10^{32}$cm$^{-2}$s$^{-1}$, while, it reaches $5.2 \times
10^{32}$cm$^{-2}$s$^{-1}$ during this winter running on the
$\psi(3770)$ peak.

During 2009, BESIII acquired a sample of 106 M $\psi'$ events, or
four times the CLEOc sample, and about 226 M $J/\psi$ events, or
about four times the BESII $J/\psi$ sample.  Results in this paper
are based on these data sets. In 2010, about 910 $pb^{-1}$
integrated luminosity was collected on the peak of $\psi(3770)$. Now
BESIII is still taking data on the peak of $\psi(3770)$, we expect
another 1.5 $pb^{-1}$ luminosity will be ready this summer.

We report on the charm physics potential at the BESIII experiment
which will make significant contribution to quark flavor physics
this decade. The charm physics program includes studies of leptonic,
semileptonic, hadronic charm decays, and tests for physics beyond
the Standard Model. High precision charm data will enable us to
validate forthcoming Lattice QCD calculations at the few percent
level. These can then be used to make precise measurements of CKM
elements, $V_{cd}$, $V_{cs}$, $V_{ub}$, $V_{cb}$ and $V_{ts}$, which
are useful to improve the accuracy of test of the CKM
unitarity~\cite{bes3book,haibonpb2006}.

\section{Results for charmonium physics}

In 2005, CLEOc~\cite{cleo} reported a measurement of the mass of the
$P$-wave charmonium spin-singlet state $h_c$ in $e^+ e^- \to \psi'
\to \pi^0 h_c,$ $h_c \to \gamma \eta_c$, in which they used both
inclusive and exclusive $\eta_c$ decay events.  In 2008, they
repeated their analysis with 25 M $\psi'$ events~\cite{cleohc}.
Combining results, they obtained $m(h_c)_{AVG} = 3525.2 \pm 0.18 \pm
0.12$ MeV/c$^2$~\cite{cleohc}. A precise determination of the mass
is important to learn about the hyperfine (spin-spin) interaction of
$P$ wave states~\cite{kuang}.  Using the spin weighted centroid of
the $^3P_J$ states, $<m(^3P_J)>$, to represent $m(^3P_J)$, they
obtained $\Delta m_{hf}(1P) =~<m(^3P_J)> - m(^1P_1) = +0.08 \pm 0.18
\pm 0.12$ MeV/c$^2$.  This is consistent with the lowest order
expectation of zero.

The $h_c$ state is also studied at BESIII with 106~M $\psi(2S)$
events accumulated in 2009~\cite{bes3hc}. Clear signals are observed
for $\psi(2S)\to \pi^0 h_c$ with and without the subsequent
radiative decay $h_c\to \gamma\eta_c$. The absolute branching ratios
$\mathcal{B}(\psi(2S) \to \pi^0 h_c) = (8.4 \pm 1.3 \pm 1.0) \times
10^{-4}$ and $\mathcal{B}(h_c \to \gamma \eta_c) = (54.3 \pm 6.7 \pm
5.2)\%$ are determined for the first time.
 The width for $h_c$ state is determined to be $\Gamma(h_c)<1.44$~MeV at the 90\% C.L..
 Measurements of $M(h_c) =
3525.40 \pm 0.13 \pm 0.18$~MeV and $\mathcal{B}(\psi' \rightarrow
\pi^0 h_c) \times \mathcal{B}(h_c \rightarrow \gamma \eta_c) = (4.58
\pm 0.40 \pm 0.50) \times 10^{-4}$ are consistent with previous
results by CLEOc~\cite{cleo}.

BESIII has studied $\psi' \to \gamma \chi_{cJ}, \chi_{cJ} \to \pi^0
\pi^0$ and $\eta \eta$, where $\pi^0$ and $\eta$ decay to $\gamma
\gamma$~\cite{bes3pi0pi0}. The branching fractions are measured to
be $\mathcal{B}(\chi_{c0} \rightarrow \pi^0\pi^0) =
(3.23\pm0.03\pm0.23\pm0.14)\times 10^{-3}$ and
$\mathcal{B}(\chi_{c2} \rightarrow \pi^0\pi^0) =
(0.88\pm0.02\pm0.06\pm0.04)\times 10^{-3}$ [$\mathcal{B}(\chi_{c0}
\rightarrow \eta\eta) = (3.44\pm0.10\pm0.24\pm0.20)\times 10^{-3}$
and $\mathcal{B}(\chi_{c2} \rightarrow \eta\eta) =
(0.65\pm0.04\pm0.05\pm0.03)\times 10^{-3}$], respectively. These
results are consistent with CLEOc measurements within
error~\cite{CLEOcpi0pi0}. Improved measurements will allow
refinement of theory~\cite{bolz}.

\section{Light hadron spectroscopy in charmonium decays}

With 106~M $\psi(2S)$ and 226~M $J/\psi$ events, BESIII
experiment~\cite{huangyp} confirmed the existence of the threshold
enhancement in the $p\bar{p}$ invariant mass in $J/\psi\to \gamma
p\bar{p}$ decay, with mass agrees with the BESII
measurement~\cite{bes2ppbar}, and the width less than 8~MeV at the
90\% C.L.. We have not observed any enhancement in the decay of
$\psi(2S)\rightarrow \gamma p\bar{p}$ at both BESII and BESIII
experiments.

In an analysis of the $J/\psi\to \gamma \eta^\prime \pi^+\pi^-$,
three resonances are observed~\cite{3etappipi}. The $X(1835)$ has a
mass of $1838\pm 3$~MeV, in good agreement with BESII
result~\cite{2etappipi}, and a width of $180\pm 9$~MeV, which is
larger than the BESII measurement. The two new resonances are
observed at 2124~MeV and 2371~MeV, respectively, with widthes about
100~MeV. The so called $X(1835)$ together with the two new
resonances are possibly the excited or higher excited $\eta$ or
$\eta^{\prime}$ states~\cite{bugg,zhu}, the candidate for glueballs,
the $p\bar{p}$ molecular states, and so on. A partial wave analysis
is needed for further understanding of these new states.

In the $\eta\pi^+\pi^-$ invariant mass recoiling against an $\omega$
in $J/\psi \rightarrow \omega \eta\pi^+\pi^-$ decay, besides the
observations of the known $f_1(1285)$ and $\eta(1405)$, a state at
$1873\pm 11$~MeV with a width of $82\pm 19$~MeV is oberserved, it
could be the hadronic production of the $X(1835)$ observed in
$\eta^\prime \pi^+\pi^-$ mode, although the mass difference is
large. Again, a partial wave analysis is urgent to be implemented in
order to understand the property of the new observation.

\section{Potential of charm physics at BESIII}

Many of the measurements related to charm decays have been done by
other experiments such as BESII and CLEO-c, and many are also
accessible to the B-factory experiments. What are BESIII's
advantages to running at the open charm threshold?

BESIII will not be able to compete both BABAR and Belle in
statistics on charm physics, especially on the rare and forbidden
decays of charm mesons. However, data taken at charm threshold still
have powerful advantages over the data at $\Upsilon(4S)$, which we
list here~\cite{gibbons}: 1) Charm events produced at threshold are
extremely clean; 2) The measurements of absolute branching fraction
can be made by using double tag events; 3) Signal/Background is
optimum at threshold; 4) Neutrino reconstruction is clean; 5)
Quantum coherence allow simple~\cite{gronau} and
complex~\cite{asner} methods to measure the neutral $D$ meson mixing
parameters and strong phase difference~\cite{liyang},  and to check
for direct $CP$ violation.

For charm physics at BESIII, the first physics results will be the
measurements of the leptonic and semileptonic decays of charm
mesons. Measurements of the leptonic decays at BESIII will benefit
from the fully tagged $D^+$ and $D^+_S$ decays available at the
$\psi(3770)$ and at $\sqrt{s} \sim 4170$ MeV or $\sim 4017$
MeV~\cite{new_scan}.  The leptonic decay rates for $D^+$ and $D^+_S$
has been measured with a precision of 4.3\% and 2.0 \% with the
final data from CLEOc. It should be noted that the $D^+ \rightarrow
\tau^+ \nu$ decay is reported by CLEOc with upper limit of $1.2
\times 10^{-3}$ at 90\% C.L.~\cite{taunu}. At BESIII, with 4 times
(about 4 fb$^{-1}$ ) of the CLEOc's luminosity, significant gains on
these measurements will be made if the systematic errors remain the
same. This will allow the validation of theoretical calculations of
the decay constants at the 1-2\% level. The neutral D mixing and
$CP$ violation in charm sector using quantum correlation are all
statistics-starved at CLEOc, improvement will be made at BESIII
experiment.

\section{Summary}

Results have been presented based on BESIII samples of 106 M $\psi'$
and about 226 $J/\psi$ events.  Many more results are to be expected
in the future.  In addition, BESIII acquired nearly $1 fb^{-1}$ of
data at the $\psi(3770)$ resonance in 2010, including approximately
$75 pb^{-1}$ of scan data around the $\psi(3770)$ peak. BESIII is
still running at the $\psi(3770)$ peak in 2011. We expect another
1.5 fb$^{-1}$ luminosity will be collected next summer.  Decays of
the $\psi(3770)$ produce quantum correlated $D \bar{D}$ pairs, which
are ideal for mixing and CP violation studies, as well measurements
of absolute branching fractions and studies of semi-leptonic decays.
This sample allows BESIII to begin their charm physics program.

\end{document}